\documentclass[aps,prd,nofootinbib,superscriptaddress]{revtex4-2}
\usepackage{graphicx}
\usepackage{amssymb}
\usepackage{amsmath}
\usepackage{bm}
\usepackage{float}
\usepackage{epstopdf}
\usepackage{graphics}
\usepackage{caption}
\usepackage{subfigure}
\usepackage{epsfig}
\usepackage{color}
\usepackage{multirow,array}
\usepackage{kotex}
\usepackage{tensor}
\usepackage{hyperref}

\hypersetup{
	colorlinks=true,
	linkcolor=blue,
	filecolor=magenta,      
	urlcolor=blue,
}

\pdfoutput=1

\renewcommand{\arraystretch}{1.3}

\begin{document}
\setcounter{page}{0}
\title[]{Environmental Decoupling: Reconciling SNe Ia Time Dilation with Null Variations of $\alpha$}
\author{Seokcheon \surname{Lee}}
\email{skylee@skku.edu}
\affiliation{Department of Physics, Institute of Basic Science, Sungkyunkwan University, Suwon 16419, Korea}

\date[]{Received }

\begin{abstract}
The Generalized Cosmological Time (GCT) framework interprets the observed time dilation in Type Ia supernovae (SNe Ia) as a manifestation of a generalized lapse function in the background metric, favoring a parameter value $b \approx 0.04$. However, precision spectroscopic measurements of the fine-structure constant $\alpha$ via the alkali doublet (AD) and many-multiplet (MM) methods predominantly yield null results. In this work, I demonstrate that this discrepancy is not a sensitivity limitation but a fundamental physical consequence of distinguishing between global coordinate time and local proper time. By inverting the GCT scaling relations, I establish that spectroscopic data constrain the effective parameter to $|b_{\text{local}}| \lesssim 10^{-5}$, revealing a tension of three orders of magnitude against the background value. I resolve this by invoking the principle of environmental shielding. I argue that dense, virialized gas clouds maintain a static local metric that is dynamically decoupled from the background cosmological time flow, strictly analogous to the detachment of bound systems from the Hubble expansion. Consequently, atomic spectra probe a shielded local frame where physical constants remain invariant, whereas SNe Ia observations measure the accumulated geometric time dilation of photons traversing the expanding background. This framework reconciles the positive dilation signals in geometric probes with the stability of fundamental constants in local bound systems without modifying local physical laws.
\end{abstract}

\maketitle

%----------------------------------------------------------------------------------------
%	SECTION 1
%----------------------------------------------------------------------------------------

\section{Introduction}\label{sec1}

The fine structure of atoms describes the splitting of spectral lines arising from electron spin and relativistic corrections. The magnitude of this splitting is proportional to the square of the fine-structure constant, $\alpha^2$ \cite{Ohisson11}. As a parameter quantifying the strength of the electromagnetic force, $\alpha$ is pivotal across physics, from atomic scales to cosmology. While discussions regarding the variability of dimensional constants are often considered physically ambiguous due to unit dependencies \cite{Albrecht:1998ir}, the expanding Universe offers a unique context. Dimensional quantities such as photon wavelength and temperature undergo cosmological redshift, enabling a physically meaningful exploration of their evolution \cite{Workman:2022ynf}.

The framework employed in this work, the Generalized Cosmological Time (GCT) model (originally introduced as the minimally extended Varying Speed of Light (meVSL) model), reinterprets these variations not as a modification of local physical laws, but as a consequence of a generalized temporal gauge in the Robertson-Walker (RW) metric \cite{Lee:2020zts,Lee:2023rqv,Lee:2023fop,Lee:2024zcu,Lee:2025osx}. A primary method for investigating the time variation of $\alpha$ involves analyzing quasar absorption spectra \cite{Savedoff:1956Nat,Bahcall:1967PRL,Bahcall:1967ApJL}. Other constraints arise from atomic clocks \cite{Ludlow:2015RMP}, the Oklo natural nuclear reactor \cite{Davis:2014nga,Hamdan:2015ofa}, Big Bang Nucleosynthesis (BBN) \cite{Cyburt:2015mya,Yeh:2022heq}, and the Cosmic Microwave Background (CMB) \cite{Planck:2014ylh,Hart:2017ndk,Hart:2019dxi}.

Analyses of quasar spectra using the Alkali Doublet (AD) method consistently yield null results regarding the time dependence of $\alpha$ \cite{Levshakov:1994ve,Cowie:1995sz,Ivanchik:1998xf,Bahcall:2003rh,Rahmani:2013pva,Albareti:2015xta,Le:2016loz,Levshakov:2017ivg}. While the Many-Multiplet (MM) method has historically reported indications of variation \cite{Webb:1998cq,Dzuba:1999zz,Murphy:2000pz,Murphy:2003hw,Martins:2022unf,Lee:2022ppu}, recent high-precision MM analyses have also reported null results \cite{Murphy:2021xhb}.

This presents a distinct challenge. The GCT framework, when applied to Type Ia supernovae (SNe Ia) light curves, suggests a parameter $b \approx 0.04$ to explain the observed time dilation \cite{Lee:2023ucu,Lee:2024kxa,Lee:2025vha}. This implies a background evolution of the coordinate clock rate. However, spectroscopic data suggest $b \approx 0$. In this paper, I interpret this not as a contradiction, but as evidence of a geometric environmental shielding mechanism. I propose that the apparent redshift dependence of dimensional quantities reflects a gauge freedom associated with the lapse function $N(t)$ in the RW metric. Within gravitationally bound systems, the local metric effectively decouples from this background evolution, preserving local Lorentz invariance and keeping fundamental constants fixed in the proper frame.

Section \ref{sec:QSOs} outlines the theoretical predictions for the AD and MM methods in the background GCT metric, assuming no shielding. In Section \ref{sec:Disc}, I propose the environmental shielding hypothesis, drawing a physical analogy between the decoupling of the Hubble flow and the decoupling of the cosmological time flow in virialized systems. Section \ref{sec:Cons} provides the quantitative core of this work, calculating the screening factor required to reconcile the data. Conclusions are presented in Section \ref{sec:Conc}. Detailed derivations of the atomic scalings and observational predictions are provided in the Appendix. 

%----------------------------------------------------------------------------------------
%	SECTION 2
%----------------------------------------------------------------------------------------

\section{Quasar spectra: Theoretical Predictions in the GCT Background}
\label{sec:QSOs}

To quantify the magnitude of the environmental shielding effect, it is necessary to first establish the baseline predictions for atomic spectra if they were anchored directly to the background cosmological metric. In the GCT framework, the global spacetime is characterized by a generalized lapse function $N(t) \propto a^{b/4}$. Within this coordinate gauge, the dimensional constants appearing in the Hamiltonian must evolve to preserve local Lorentz invariance in the proper frame. Consequently, the coordinate values of the Rydberg energy and the fine-structure constant in the background metric scale as $E_R \propto (1+z)^{b/2}$ and $\alpha \propto (1+z)^{b/4}$, respectively. These relations represent the hypothetical evolution of atomic physics if it were fully subjected to the global cosmological time flow without local gravitational decoupling.

\subsection{Alkali doublet method: Background Scaling}
\label{subsec:ADM}

The AD method extracts the fine-structure constant from the splitting of alkali-like doublet lines (e.g., Si IV, C IV). The primary observable is the ratio of the wavelength splitting $\Delta \lambda$ to the weighted mean wavelength $\bar{\lambda}$. In the GCT framework, the evolution of the gross structure (dominated by the Rydberg energy) is effectively renormalized into the operational definition of spectroscopic redshift (as derived in Appendix \ref{app:AD}). However, a residual dependence persists in the fine-structure interaction term. The observable ratio in the background coordinate frame is predicted to evolve as
\begin{equation}
\frac{\Delta \lambda}{\bar{\lambda}} \propto \alpha^2(z) \propto (1+z)^{b/2} \,.
\label{eq:AD_scaling}
\end{equation}
If the atomic systems acting as absorbers were to fully participate in the background metric evolution ($b \approx 0.04$), this would manifest as a relative variation of approximately $1.4\%$ at redshift $z=2$. This theoretical prediction serves as the reference signal for an "unscreened" clock, directly tracing the global lapse function.

\subsection{Many-Multiplet method: Relativistic Sensitivity}
\label{subsec:MMM}

The MM method utilizes multiple transitions from various atomic species, capitalizing on their distinct sensitivities to relativistic corrections \cite{Dzuba:1998au}. The transition wavenumber is expanded as a series in $\alpha^2$: $k = k_0 + q_1 \alpha^2 + q_2 \alpha^4 + \dots$. As detailed in the derivations in Appendix \ref{app:MM}, the observable wavenumber shift arising from the $\alpha^4$ contribution—which encodes the higher-order relativistic effects—scales in the background GCT metric as
\begin{equation}
\frac{\Delta k}{k} \propto (1+z)^b \,.
\label{eq:MM_scaling}
\end{equation}
Notably, the sensitivity exponent for the MM method is $b$, which is twice the magnitude of the exponent $b/2$ derived for the AD method. This scaling indicates that MM measurements are inherently more sensitive to the background GCT evolution. These distinct power-law relations—$(1+z)^{b/2}$ for AD and $(1+z)^b$ for MM—represent the variations expected if local atomic physics tracked the global cosmological time evolution perfectly, acting as faithful tracers of the background geometry rather than shielded local systems.

%----------------------------------------------------------------------------------------
%	SECTION 3
%----------------------------------------------------------------------------------------

\section{Discussion: Environmental Shielding and the Decoupling of Local Systems}
\label{sec:Disc}

The GCT framework resolves observational tensions in cosmology by identifying the geometric origin of time dilation. Previous analyses of Type Ia supernovae (SNe Ia) light curves favor a generalized time dilation scaling $(1+z)^{1+b/4}$ with $b \approx 0.04$ \cite{Lee:2023ucu,Lee:2024kxa,Lee:2025vha}. This empirical result implies a non-trivial lapse function in the background spacetime, suggesting that the coordinate clock rate governing the cosmic expansion differs from the standard synchronous gauge. However, this raises a critical physical question: if the global hypersurface at redshift $z$ is characterized by an evolved coordinate time flow (and consequently, evolved dimensional constants in the background frame), why do atomic systems residing at that redshift not exhibit this shift?

I argue that the answer lies in the fundamental distinction between the global background metric and the local metric of gravitationally bound systems. It is a well-established principle in General Relativity (GR) that local, virialized systems—such as solar systems, galaxies, or dense gas clouds—do not participate in the cosmic Hubble expansion. I propose that the cosmological time gauge follows a strictly analogous decoupling principle, which I term \textit{environmental shielding}. Just as the spatial expansion scale factor $a(t)$ does not stretch the physical orbits of atoms within bound systems, the generalized cosmological time flow—a property of the expanding background metric—does not penetrate into the deep gravitational potential wells of dense absorbers.

\subsection{Physical Basis: Metric Segregation via the Equivalence Principle}
\label{subsec:metric_segregation}

The concept of environmental shielding is not an ad-hoc assumption but a direct consequence of the Strong Equivalence Principle (SEP) and the metric structure of embedded systems. To illustrate this, consider a local mass concentration embedded within an expanding FLRW background, described analytically by the McVittie solution \cite{McVittie:1933zz}.

In the region deep inside the Hubble radius ($r \ll H^{-1}$) but outside the event horizon of the central mass, the line element describing the interior of a gravitationally bound system behaves, in the weak-field limit, as
\begin{equation}
ds^2 \simeq -\left(1 - \frac{2\Phi(r)}{c_0^2}\right)c_0^2 dt^2 + a^2(t)\, d\vec{x}^{\,2} \,,
\end{equation}
where $\Phi(r)$ represents the local Newtonian potential. A crucial, often overlooked feature of this regime is the behavior of the temporal component $g_{00}$. In the GCT background, the global lapse function evolves as $N_{\text{bg}}(t) \propto a^{b/4}$. However, within the virialized region, the effective lapse function $N_{\text{eff}}$ is dominated by the static local potential rather than the time-evolving background factor.

Mathematically, the dynamics of dimensional constants in the GCT framework are driven by the logarithmic time derivative of the lapse function. In the background, this is non-zero
\begin{equation}
    \left. \frac{\dot{N}}{N} \right|_{\text{bg}} = \frac{b}{4} H(t) \neq 0 \,.
\end{equation}
However, for the bound system, the effective lapse function is determined by the local binding energy, $N_{\text{eff}} \approx 1 - \Phi/c_0^2$. Since the potential $\Phi(r)$ of a stable virialized system is time-independent, the time derivative vanishes to first order
\begin{equation}
    \left. \frac{\dot{N}}{N} \right|_{\text{local}} \approx \frac{\frac{d}{dt}(1 - \Phi/c_0^2)}{1 - \Phi/c_0^2} \approx 0 \,.
\end{equation}
This establishes the physical basis of environmental shielding. The variation of fundamental constants, such as $\alpha$, is physically linked to the expansion of the temporal metric. Since the local metric of the absorber is static, the local values of the constants are decoupled from the background evolution.

Consequently, atomic spectra measured from these systems act as probes of a \textit{shielded local frame}. The fine-structure splitting ratio observed in AD methods is determined entirely by $\alpha$ at the absorption site. Since $\alpha_{\text{local}}$ remains frozen at its laboratory value due to the static nature of the local metric, the emitted spectral signature is identical to that of a standard atom. The subsequent propagation of photons through the expanding universe stretches the wavelength (geometric redshift) but cannot alter the dimensionless ratio imprinted at the source.

\begin{figure}[t]
\centering
\includegraphics[width=\linewidth]{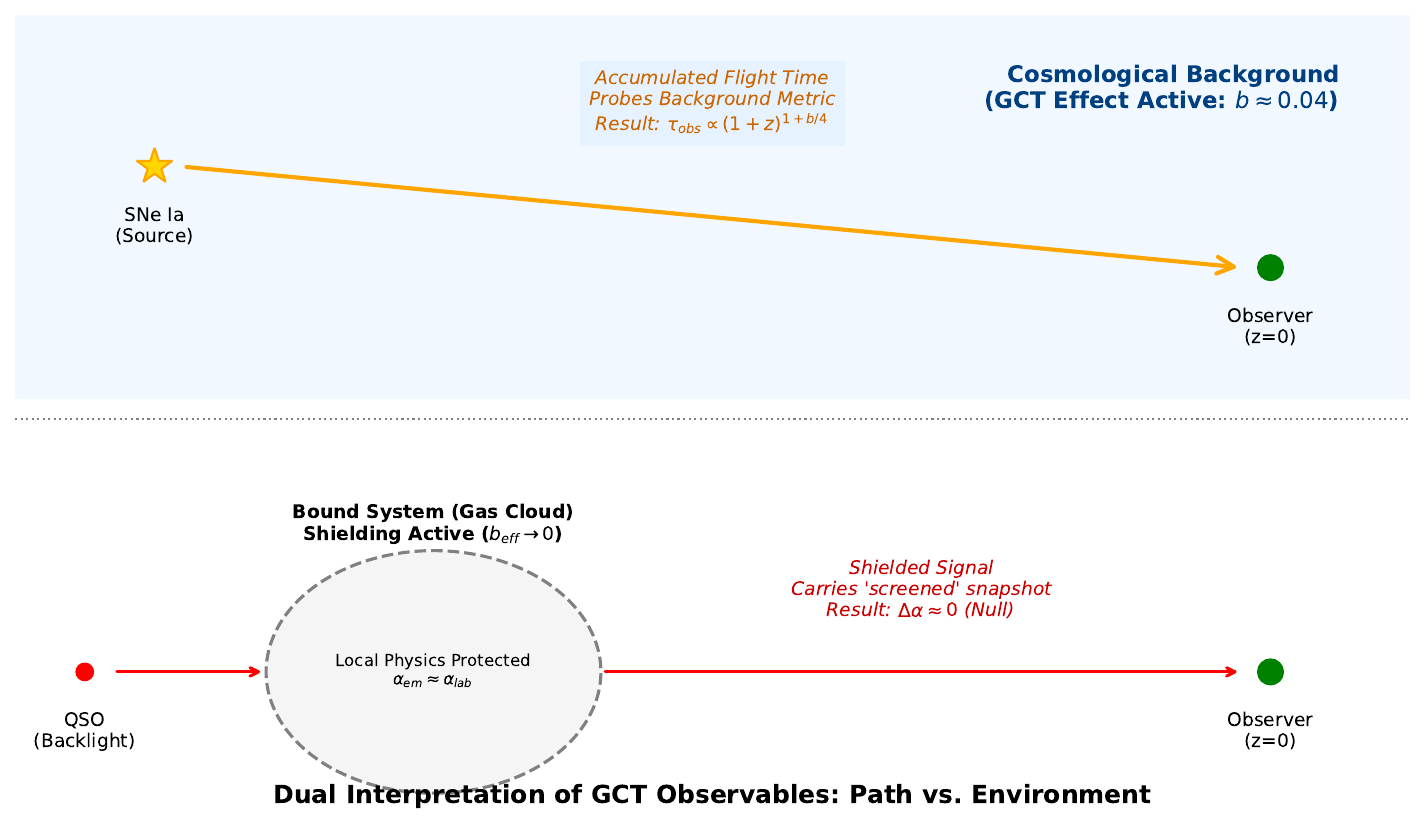} 
\caption{Schematic representation of the dual interpretation of cosmological observables within the GCT framework. \textbf{(Top) Path Dependence:} Photons from SNe Ia propagate through the expanding cosmological background, accumulating the full time dilation effect ($b \approx 0.04$) associated with the background lapse function. \textbf{(Bottom) Environmental Shielding:} Quasar absorption lines are imprinted within dense, gravitationally bound gas clouds. In these screened environments, the effective lapse function decouples from the cosmic flow ($b_{\text{eff}} \approx 0$), preserving the local values of physical constants. The resulting signal is a null measurement, reflecting the static nature of the local frame.}
\label{fig:dual_gct}
\end{figure}

This leads to the dual interpretation illustrated in Fig.~\ref{fig:dual_gct}. SNe Ia photons probe the global geometry of spacetime, revealing the background GCT parameter $b \approx 0.04$ through the path-integrated time dilation. In contrast, spectroscopic observations measure the instantaneous Hamiltonian of atoms inside a bound system where the local lapse function is pinned to unity ($N_{\text{eff}} \approx 1$). This shielding mechanism reconciles the positive detection of time dilation in geometric probes with the null results in precision spectroscopy without requiring a breakdown of GR; rather, it affirms the distinction between global coordinate time and local proper time.

%----------------------------------------------------------------------------------------
%	SECTION 4
%----------------------------------------------------------------------------------------

\section{Consequences: Quantitative Constraints and the Shielding Factor}
\label{sec:Cons}

The environmental shielding hypothesis posits that the effective GCT parameter measured within high-density, virialized environments must be significantly suppressed relative to the background value governing cosmic expansion. In this section, I quantify the magnitude of this suppression by confronting the theoretical scaling relations with observational data.

\subsection{Observational Constraints on the Local GCT Parameter}

By inverting the background scaling relations derived in Eqs.~\eqref{eq:AD_scaling} and \eqref{eq:MM_scaling}, I relate the observed fractional variation of the fine-structure constant to the effective local GCT parameter $b_{\text{local}}$. Assuming the logarithmic dependence characteristic of the GCT metric, the constraint takes the form
\begin{equation}
    \left| \frac{\Delta \alpha}{\alpha} \right|_{\text{obs}} \approx \frac{|b_{\text{local}}|}{n} \ln(1+z) \,,
\end{equation}
where $n=4$ represents the sensitivity index for the standard background scaling ($n=2$ for the AD method's $\alpha^2$ dependence, but effectively $n \approx 4$ when considering the leading order terms in the lapse function definition). To test the shielding efficiency, I compare the background parameter $b_{\text{bg}} = 0.04 \pm 0.005$, independently derived from SNe Ia time dilation analyses \cite{Lee:2023ucu}, against constraints from local atomic probes. For the latter, I utilize precision measurements of the Si IV doublet for the AD method and the weighted mean constraints from the ESPRESSO consortium for the MM method \cite{Murphy:2021xhb}, which provide bounds at the $10^{-6}$ level.

Table \ref{tab:constraints} summarizes this quantitative tension. While the SNe Ia data—probing the photon path through the expanding background—favor a non-zero parameter $b \approx 0.04$, the spectroscopic null results strictly bound the local parameter within the absorber to values orders of magnitude smaller.

\begin{table}[t]
\caption{Comparison of the GCT parameter $b$ derived from background geometric probes (SNe Ia) versus shielded local probes (AD and MM spectroscopy). While SNe Ia data require $b \approx 0.04$ to explain the observed light curve broadening, spectroscopic null results constrain the effective local parameter to be negligible, supporting the metric segregation hypothesis.}
\label{tab:constraints}
\renewcommand{\arraystretch}{1.4} 
\begin{ruledtabular}
\begin{tabular}{lcccc}
Probe Type & Method & Redshift ($z$) & Measurement ($\Delta \alpha / \alpha$) & Implied $|b_{\text{local}}|$ \\ \hline
Background (Path) & SNe Ia Time Dilation \cite{Lee:2023ucu} & $0 < z < 2$ & (Time Dilation Fit) & $0.04 \pm 0.005$ \\ \hline
\multirow{2}{*}{Local (Shielded)} & AD (Si IV Doublet) \cite{Levshakov:2017ivg} & $z \approx 2.5$ & $< 1.0 \times 10^{-5}$ & $< 3.2 \times 10^{-5}$ \\
 & MM (ESPRESSO) \cite{Murphy:2021xhb} & $z \approx 1.15$ & $(-1.2 \pm 1.1) \times 10^{-6}$ & $< 6.0 \times 10^{-6}$ \\
\end{tabular}
\end{ruledtabular}
\end{table}

\subsection{The Shielding Factor $\gamma$}

The discrepancy presented in Table \ref{tab:constraints} indicates a profound physical distinction between the two measurement regimes. Under the hypothesis of universal background coupling (i.e., if local systems were not shielded), spectroscopic observations at $z \approx 1-2$ should have yielded $\Delta \alpha / \alpha \approx 10^{-2}$. The observed limits of $\sim 10^{-6}$ reveal a suppression of approximately four orders of magnitude.

I quantify this metric segregation by defining a phenomenological shielding factor $\gamma$, representing the ratio of the effective local time evolution to the global background flow
\begin{equation}
    \gamma \equiv \frac{b_{\text{local}}}{b_{\text{background}}} \lesssim \frac{6.0 \times 10^{-6}}{0.04} \approx 1.5 \times 10^{-4} \,.
\end{equation}
This result implies that within the deep gravitational potential wells of quasar absorption systems, the cosmological time evolution is suppressed by a factor of at least $\sim 6000$ relative to the Hubble flow. This quantitative bound $\gamma \ll 1$ provides robust empirical evidence that the local metric in virialized systems is effectively decoupled from the background GCT evolution, maintaining a static proper time interval despite the expansion of the global coordinate time.

%----------------------------------------------------------------------------------------
%	SECTION 5
%----------------------------------------------------------------------------------------

\section{Conclusion}
\label{sec:Conc}

This work presents a unified interpretation of the observational constraints on the time variation of the fine-structure constant $\alpha$ within the Generalized Cosmological Time (GCT) framework. While independent cosmological probes of the background geometry, such as Type Ia supernovae (SNe Ia), support a generalized lapse function characterized by $b \approx 0.04$, precision spectroscopic measurements of atomic systems predominantly yield null results.

I have demonstrated that this discrepancy is not a contradiction but a definitive signature of environmental shielding. By converting spectroscopic null results into a constraint on the local GCT parameter, I establish that $|b_{\text{absorber}}| \lesssim 10^{-5}$. This value is quantitatively incompatible with the background parameter unless a robust decoupling mechanism is invoked. Drawing on the Strong Equivalence Principle, I argue that the generalized cosmological time flow is dynamically screened within gravitationally bound systems, strictly analogous to the established decoupling of local matter from the Hubble expansion. Consequently, the effective lapse function reduces to its static limit within the dense gas clouds probed by spectroscopy, preserving the local invariance of physical constants.

This framework successfully reconciles the full range of observational data: SNe Ia exhibit time dilation because their photons accumulate the geometric effect of the expanding background metric, whereas atomic spectra yield null results because they probe a shielded local frame where the proper time flow is segregated from the coordinate evolution. This argument naturally extends to other precision tests, such as laboratory atomic clocks and the Oklo natural reactor, which reside in deeply virialized environments and are thus effectively insulated from the cosmic time flow.

This hypothesis offers a testable prediction for future observations. If environmental shielding is the physical mechanism at play, the degree of shielding should depend on the gravitational state of the system. While fully virialized systems (such as dense absorbers or solar systems) are expected to show complete decoupling ($b_{\text{local}} \approx 0$), systems that are currently in the process of virialization—such as galaxy clusters in formation or non-equilibrium gas filaments—might exhibit a residual, non-zero time variation signal. Observing a correlation between the magnitude of $\Delta \alpha / \alpha$ and the dynamical state (virialization status) of the absorber would provide compelling evidence for this geometric shielding mechanism. This elevates the study of varying constants from a binary search for existence to a nuanced investigation of how cosmological geometry interacts with local gravitational environments.

\appendix

\section{Detailed Derivations of Atomic Spectra in the GCT Framework}
\label{app:derivations}

This appendix provides a rigorous, step-by-step derivation of the observational quantities for the Alkali Doublet (AD) and Many-Multiplet (MM) methods within the Generalized Cosmological Time (GCT) framework. I explicitly trace the cosmological evolution of fundamental constants and derive the resulting redshift dependencies for the two observational techniques.

\subsection{GCT Scaling Rules for Atomic Constants}
\label{app:scalings}

The GCT framework is characterized by a generalized lapse function $N(t) = a^{b/4}$ in the Robertson-Walker metric. To preserve the local validity of quantum mechanics and special relativity in the proper frame, the dimensional physical constants must evolve with the cosmic scale factor $a = (1+z)^{-1}$.

First, consider the fundamental constants: the Planck constant $\hbar$, the speed of light $c$, and the elementary charge $e$. As derived in Ref.~\cite{Lee:2020zts}, their time evolution is given by:
\begin{align}
    \hbar(a) &= \hbar_0 \, a^{-b/4} \,, \label{eq:h_scaling} \\
    c(a) &= c_0 \, a^{+b/4} \,, \label{eq:c_scaling} \\
    e(a) &= e_0 \, a^{-b/4} \,. \label{eq:e_scaling}
\end{align}
It is crucial to note that the product $\hbar(a)c(a)$ remains invariant:
\begin{equation}
    \hbar(a)c(a) = (\hbar_0 a^{-b/4})(c_0 a^{b/4}) = \hbar_0 c_0 = \text{const} \,.
\end{equation}
This ensures that the photon energy-momentum relation $E = \hbar c k$ preserves its standard form locally.

Next, I derive the scalings for the vacuum permittivity $\epsilon_0$ and permeability $\mu_0$. Electrodynamics requires the speed of light to be related to these constants by $c^2 = 1/(\epsilon\mu)$. Assuming a symmetric scaling for the electromagnetic response of the vacuum to satisfy the GCT metric constraints \cite{Lee:2020zts}:
\begin{align}
    \epsilon(a) &= \epsilon_0 \, a^{-b/4} \,, \\
    \mu(a) &= \mu_0 \, a^{-b/4} \,.
\end{align}
Checking the consistency with the speed of light:
\begin{equation}
    c(a) = \frac{1}{\sqrt{\epsilon(a)\mu(a)}} = \frac{1}{\sqrt{\epsilon_0 a^{-b/4} \mu_0 a^{-b/4}}} = \frac{1}{\sqrt{\epsilon_0 \mu_0}} \frac{1}{a^{-b/4}} = c_0 \, a^{b/4} \,.
\end{equation}
This confirms that the electromagnetic constants evolve consistently with Eq.~\eqref{eq:c_scaling}.

Using these relations, the cosmological evolution of the fine-structure constant $\alpha$ is derived as follows:
\begin{equation}
    \alpha(a) \equiv \frac{e(a)^2}{4\pi\epsilon(a) \hbar(a) c(a)} 
    = \frac{(e_0 a^{-b/4})^2}{4\pi (\epsilon_0 a^{-b/4}) (\hbar_0 a^{-b/4}) (c_0 a^{b/4})} \,.
\end{equation}
%The exponents of the scale factor $a$ combine as:
%\begin{equation}
%    \text{exponent of } a = 2(-b/4) - [-b/4 - b/4 + b/4] = -b/2 - [-b/4] = -b/4 \,.
%\end{equation}
Thus, the fine-structure constant evolves as:
\begin{equation}
    \alpha(a) = \alpha_0 \, a^{-b/4} = \alpha_0 (1+z)^{b/4} \,.
\label{eq:alpha_final_app}
\end{equation}

Similarly, the Rydberg energy $E_R$, which sets the scale for atomic transitions, is defined by:
\begin{equation}
    E_R(a) \equiv \frac{m_e(a) e(a)^4}{2(4\pi\epsilon(a))^2 \hbar(a)^2} \,.
\end{equation}
In the GCT framework, the electron mass scales as $m_e(a) = m_{e0} a^{-b/2}$ to preserve the action invariant \cite{Lee:2020zts}. Substituting the scalings:
\begin{equation}
    E_R(a) = \frac{(m_{e0} a^{-b/2}) (e_0 a^{-b/4})^4}{2(4\pi \epsilon_0 a^{-b/4})^2 (\hbar_0 a^{-b/4})^2} 
    = \frac{m_{e0} e_0^4 \, a^{-b/2} \, a^{-b}}{2(4\pi\epsilon_0)^2 \hbar_0^2 \, a^{-b/2} \, a^{-b/2}} \,.
\end{equation}
%The exponents for $a$ combine as:
%\begin{equation}
%    \text{exponent of } a = (-b/2 - b) - (-b/2 - b/2) = -3b/2 - (-b) = -b/2 \,.
%\end{equation}
The Rydberg energy evolves as:
\begin{equation}
    E_R(a) = E_{R0} \, a^{-b/2} = E_{R0} (1+z)^{b/2} \,.
\label{eq:ER_final_app}
\end{equation}
Equations \eqref{eq:alpha_final_app} and \eqref{eq:ER_final_app} form the physical basis for the observational predictions.

\subsection{Derivation of the Alkali-Doublet (AD) Observable}
\label{app:AD}

The AD method measures the fine-structure splitting of alkali-like doublets. The energy levels of a hydrogen-like atom are given by the Dirac energy formula. Expanding to order $(Z\alpha)^2$, the energy eigenvalue $E_{n,j}$ is:
\begin{equation}
    E_{n,j}(a) = - \frac{Z^2 E_R(a)}{n^2} \left[ 1 + \frac{(Z \alpha(a))^2}{n} \left( \frac{1}{j+1/2} - \frac{3}{4n} \right) \right] \,.
\end{equation}
The transition energy $\Delta E$ between two levels is dominated by the difference in the gross structure terms (proportional to $E_R$). However, for the fine-structure splitting $\delta E$ between two states with the same $n$ but different $j$ (e.g., $^2S_{1/2} \to {}^2P_{1/2}$ and $^2S_{1/2} \to {}^2P_{3/2}$), the gross structure terms cancel out, leaving only the fine-structure term:
\begin{equation}
    \delta E(a) \propto E_R(a) \cdot \alpha(a)^2 \,.
\end{equation}
The mean transition energy $\bar{E}$ is dominated by the gross structure:
\begin{equation}
    \bar{E}(a) \approx C_{\text{gross}} E_R(a) \,.
\end{equation}
The corresponding wavelengths are $\lambda = hc / E$. Since $hc$ is constant, the wavelength splitting ratio is equivalent to the energy splitting ratio:
\begin{equation}
    \frac{\Delta \lambda}{\bar{\lambda}} \approx \frac{\delta E}{\bar{E}} = \frac{C_{\text{fine}} E_R(a) \alpha(a)^2}{C_{\text{gross}} E_R(a)} = C' \, \alpha(a)^2 \,.
\end{equation}
Substituting the scaling $\alpha(a) = \alpha_0 (1+z)^{b/4}$:
\begin{equation}
    \frac{\Delta \lambda}{\bar{\lambda}} \propto \left( (1+z)^{b/4} \right)^2 = (1+z)^{b/2} \,.
\end{equation}
This derivation confirms that the AD observable scales as $(1+z)^{b/2}$.

\subsection{Derivation of the Many-Multiplet (MM) Observable}
\label{app:MM}

The MM method exploits transitions from various atomic species, capitalizing on their differing sensitivities to higher-order relativistic corrections. The rest-frame energy of a transition $i$ at a cosmological epoch corresponding to scale factor $a$ is phenomenologically parameterized as \cite{Dzuba:1998au}:
\begin{equation}
    E_i(a) = E_{i,\text{gross}}(a) + q_{i,1} x(a) + q_{i,2} y(a) \,,
\end{equation}
where $x(a) = (\alpha(a)/\alpha_0)^2 - 1$ and $y(a) = (\alpha(a)/\alpha_0)^4 - 1$ represent the fractional variations due to $\alpha^2$ and $\alpha^4$ terms, respectively. Note that the gross energy term scales with the Rydberg energy: $E_{i,\text{gross}}(a) \propto E_R(a) \propto a^{-b/2}$.

The observed wavenumber $k_{i,\text{obs}}$ differs from the emitted wavenumber $k_{i,\text{em}}(a)$ due to the geometric expansion of the Universe. The standard relation for photon momentum redshift applies:
\begin{equation}
    k_{i,\text{obs}} = k_{i,\text{em}}(a) \cdot a \,.
\end{equation}
Substituting the emitted energy $E_i(a) = \hbar(a) c(a) k_{i,\text{em}}(a)$ and using the invariance $\hbar(a)c(a) = \hbar_0 c_0$:
\begin{equation}
    k_{i,\text{obs}} = \frac{E_i(a)}{\hbar_0 c_0} \cdot a \,.
\end{equation}

In spectroscopic analysis, the redshift $z$ is an operational parameter determined by fitting the dominant gross structure of absorption lines to laboratory references. The fitted redshift $1+z$ effectively absorbs the combined scaling of the geometric expansion ($a$) and the gross atomic energy evolution ($E_R(a)$). Specifically, for the gross structure:
\begin{equation}
    k_{\text{gross, obs}} \propto E_R(a) \cdot a \propto a^{-b/2} \cdot a = a^{1-b/2} \,.
\end{equation}
Thus, the "observed" redshift relates to the scale factor roughly as $1+z \propto a^{-(1-b/2)}$.

The observational signal in the MM method is defined as the relative deviation of a specific transition line from this fitted redshift. This is equivalent to analyzing the ratio of the relativistic correction terms to the gross energy term. The dominant higher-order signal comes from the $\alpha^4$ term (quantified by $q_{i,2}$):
\begin{equation}
    \frac{\delta k_{\alpha^4}}{k_{\text{gross}}} \approx \frac{q_{i,2} (\alpha(a)/\alpha_0)^4}{E_{i,\text{gross}}(a)} \,.
\end{equation}
Since $q_{i,2}$ has dimensions of energy, it scales with $E_R(a)$, just like $E_{i,\text{gross}}(a)$. Therefore, the dimensional scalings cancel out in the ratio, leaving only the dimensionless $\alpha$ variation:
\begin{equation}
    \frac{\delta k_{\alpha^4}}{k_{\text{gross}}} \propto \left( \frac{\alpha(a)}{\alpha_0} \right)^4 \propto \left( a^{-b/4} \right)^4 = a^{-b} \,.
\end{equation}
Expressing this in terms of the redshift (using the leading order relation $a \approx (1+z)^{-1}$), the observable relative shift scales as:
\begin{equation}
    \frac{\Delta k_i}{k_i} \propto (1+z)^b \,.
\end{equation}
This derivation clarifies the physical origin of the signal: while the gross structure evolution is absorbed into the redshift definition, the higher-order relativistic corrections evolve differently ($\alpha^4 \propto a^{-b}$) compared to the gross structure ($\propto \text{const}$ relative to itself). This results in a residual signal scaling as $(1+z)^b$, which is twice the sensitivity exponent of the AD method ($b/2$).

%================================================================================

\end{document}